\documentclass[intlimits,twoside,a4paper]{article}
\usepackage{wrapfig}
\usepackage{amssymb}
\usepackage[T2A]{fontenc}
\usepackage[cp1251]{inputenc}
\usepackage{amsmath}
\usepackage{graphicx}
\usepackage{epstopdf}

\usepackage{cmpj2}

\def\be{\begin{equation}}
\def\ee{\end{equation}}
\def\bea{\begin{eqnarray}}
\def\eea{\end{eqnarray}}

\def\gsim{\mathrel{\lower.65ex\hbox{$\mathop{\kern0pt\sim}\limits
   ^{\lower.55ex\hbox{$>$}}$}}}
\def\lsim{\mathrel{\lower.65ex\hbox{$\mathop{\kern0pt\sim}\limits
   ^{\lower.55ex\hbox{$<$}}$}}}

\DeclareMathOperator\erfc{erfc}

\issue{2015}{18}{1}{13602}
\doinumber{10.5488/CMP.18.13602}

\title[Fluid surfaces in external field]
{Application of molecular simulations: Insight into liquid bridging and jetting phenomena}
\author[I. Nezbeda \textsl{et al.}]
{
I. Nezbeda\refaddr{PU,UCHP},
J. Jirs\'{a}k\refaddr{PU},
F. Mou\v{c}ka\refaddr{PU},
W.R. Smith\refaddr{UoG}}
\addresses{
\addr{PU}
Faculty of Science,
J.E. Purkinje University, 400\,96 \'{U}st\'{\i}n. Lab., Czech Republic
\addr{UCHP} Institute of Chemical Process Fundamentals,
Academy of Sciences, 165\,02 Prague 6 -- Suchdol, Czech Republic
\addr{UoG} Department of Mathematics and Statistics,
University of Guelph, Guelph, ON N1G 2W1, Canada
}

\date{Received October 22, 2014, in final form December 20, 2014}
\authorcopyright{I. Nezbeda, J. Jirs\'{a}k, F. Mou\v{c}ka, W.R. Smith, 2015}

\begin{document}

\maketitle

\begin{abstract}

Molecular dynamics simulations have been performed on pure liquid water, aqueous solutions of sodium chloride, and polymer solutions exposed to a strong external electric field with the goal to gain molecular insight into the structural response to the field. Several simulation methodologies have been used to elucidate the molecular mechanisms of the processes leading to the formation of liquid bridges and jets (in the production of nanofibers). It is shown that in the established nanoscale structures, the molecules form a chain with their dipole moments oriented parallel to the applied field throughout the entire sample volume. The presence of ions may disturb this structure leading to its ultimate disintegration into droplets; the concentration dependence of the threshold field required to stabilize a liquid column has been determined. Conformational changes of the polymer in the jetting process have also been observed.


\pacs 68.03.Hj
\end{abstract}

\section{Introduction}

With advance in computer technology, the development of efficient simulation techniques, and the availability of commercial or freeware packages, molecular simulation methods \cite{Allen,SmitFrenkel} have become the main, and in some instances the only tool to study fluid systems at the molecular level.
A primary output of molecular simulations (and target of theories) have typically been thermophysical properties (equation of state, heat capacity, viscosity, etc.), entering macroscopic theories of the studied systems, and their relation to various kinds of underlying intermolecular interactions. However, they make it also possible to study various processes and this approach has become thus commonplace in many other fields, including mechanical engineering and material design, biology, and even medicine, often shedding light on problems that are inaccessible by other methods.
Examples include the rheology of lubricants~\cite{rheol1,rheol2} and protein self-assembly~\cite{protein}, potentially related to Alzheimer's and Parkinson's diseases. In these applications, the simulations need not necessarily yield numerical values of physical quantities, but they may help to qualitatively understand and explain the underlying molecular mechanism of the observed phenomena.

A similar type of problems whose molecular nature is not yet fully understood involves liquid surfaces exposed to strong external electric fields, their disturbance and subsequent development. These problems are encountered in a variety of fields, e.g., physics of the atmosphere (ice formation in supercooled water), biology (transport through biological cell membranes), and mineralogy (reactivity at the mineral--water interface).
Interesting examples of processes in which the external field is the driving force are the floating water bridge (FWB), whose application potential has not been explored yet~\cite{FuchsAppl}, and the production of nanofibers using needleless electrospinning technology \cite{ElspinRev}.

\begin{figure}[!t]
  \begin{center}
    \includegraphics[width=0.8\textwidth]{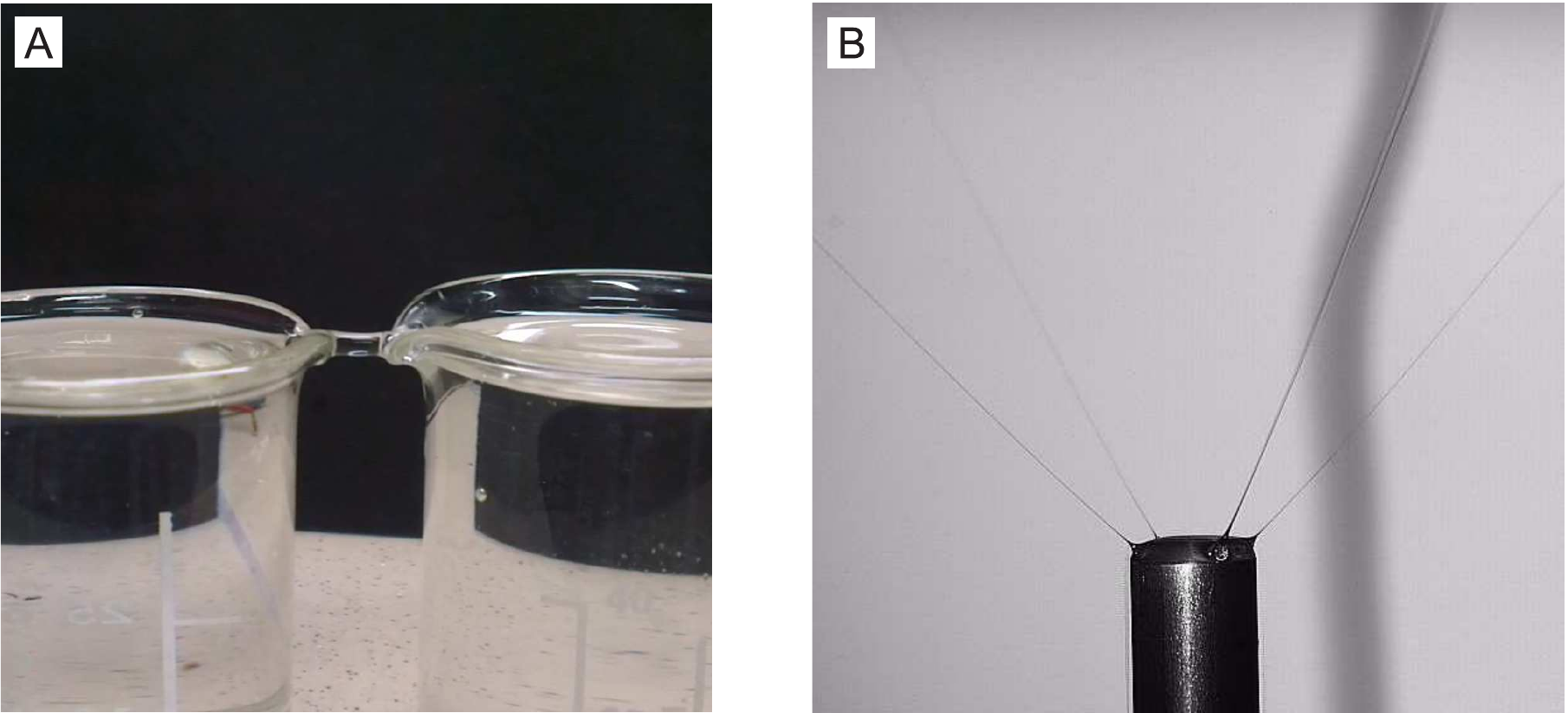}
  \end{center}
  \caption{Photographs of (A) the floating water bridge established between two beakers and (B) jets emanating from the macroscopic drop in electrospinning.}
  \label{fig_pictures}
\end{figure}

The macroscopic floating bridge arises between two beakers filled with a liquid (not necessarily water) and with immersed electrodes~\cite{FWB}, see figure~\ref{fig_pictures}~(A). The beakers are initially in contact and a free hanging structure is formed when they are then slowly pulled apart. A nanoscale analogue to the floating bridge (a nanobridge) is observed between the tip and sample surface in atomic force microscopy (AFM) \cite{AFM}.
Several scattering experiments using different techniques have been used to elucidate the microscopic mechanism which holds the FWB together but their results were inconclusive or even contradicting \cite{scatter1,scatter2,scatter3,Skinner}. And, as for molecular simulation studies, we are aware of only two attempts. Skinner et al.~\cite{Skinner} used the simulation setup not fully representing the FWB but only its bulk-like interior and did not find any preferential orientation of the water molecules in the bridge.
Cramer et al.~\cite{Cramer} studied a drop attached to a surface and observed its deformation into a column-like structure, though never fully developing a bridge. Their main result was the finding that whatever strength of the field was used, the hydrogen bond network of the water molecules remained practically intact.

\begin{figure}[!b]
  \begin{center}
    \includegraphics[width=0.65\textwidth]{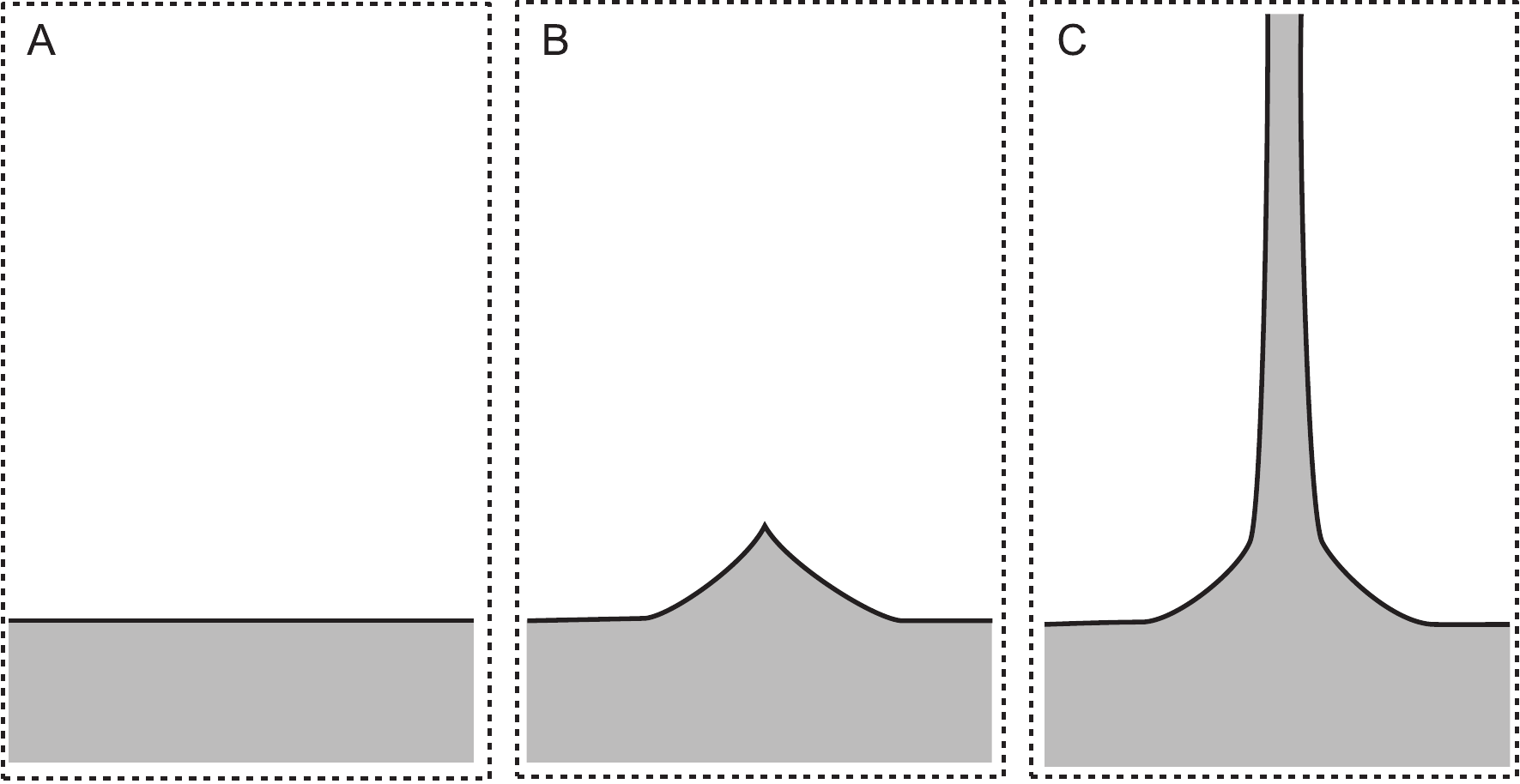}
  \end{center}
  \caption{Schematic representation of the formation of the jet: (A) the initial flat surface; (B) formation of the cone; and (C) eruption of the jet. }
  \label{fig_scheme}
\end{figure}

In needleless electrospinning, the external field destabilizes the surface of the liquid. It gives rise to the formation of a Taylor cone which ultimately looses stability and an eruption of a jet is observed; these stages are schematically shown in figure~\ref{fig_scheme}. The photograph in figure~\ref{fig_pictures}~(B) depicts the experimental observation of jetting. After flying through a short linear section, the jet undergoes the so-called whipping stage and the resulting fiber is then collected at the other electrode. With the exception of a recent paper~\cite{JMN_1}, we are not aware of any other atomistic simulation study dealing directly with the above described process. Other available simulations related to this process focus on the structure formation of polymer chains~\cite{24}, the effect of solvent on the produced nanofibers~\cite{25}, and on spraying ionic liquids~\cite{26}.

In a recent paper~\cite{JMN_1} we made the first attempt to apply molecular simulations to the electrospinning process to examine (i) to what extent simulation can be applied to such a dynamic process and (ii) what results it can provide. We used three different methodologies, all of which yielded qualitatively the same results. We have continued in this research in which, in addition to the qualitative pictures, the gain of also some quantitative characterizations has been attempted and the obtained results are reported in this paper. We consider both the solution of sodium chloride and a model polymer solution, specifically the solution of polyethylene glycol (PEG). For completeness and comparison, also pure water has been included in this study.
In the next section we provide the main technical details of the simulations, and in section~3 we present the obtained results. These include the stability analysis in the dependence on the strength of the field, an analysis of the orientational arrangement of the water molecule dipoles and the effect of the presence of ions thereupon in dependence on the electrolyte concentration, changes in the polymer conformation during the process, and relevant kinetic properties.

\section{Methodology}

For molecular dynamics simulations reported in this paper, we used the GROMACS package (versions 4.5 and 4.6)~\cite{GROMACS}. A leap-frog integrator with time steps ranging from 1 to 2~fs was employed and the particle-mesh Ewald method~\cite{PME} with cubic interpolation and tinfoil boundary conditions were used to treat electrostatic forces whenever periodic boundary conditions were applied. The technical parameters of the Ewald summation (cutoff, $r_\mathrm{c}$,  separation parameter, $\alpha$, grid spacing, etc.) were being adjusted for different setups individually to optimize computational cost while maintaining reasonable accuracy. The relative strength of the direct potential at the cutoff (GROMACS \texttt{ewald-rtol} parameter), i.e., $\erfc(\alpha r_\mathrm{c})$, was always set to the value 10$^{-5}$.

To keep contact with the previous study~\cite{JMN_1}, for simulations with pure water and sodium chloride solutions, we used the TIP3P water model~\cite{TIP3P} and the associated Joung and Cheatham~\cite{Joung} ionic force fields. The solution of polyethylene glycol (PEG) was modelled by the GROMOS~53A6$_{\textrm{OXY}+\textrm{D}}$~\cite{PEG} force field with SPC water.

In order to approach the proper scale of the investigated processes as closely as possible, and to make simultaneously the simulation feasible using the available computing facilities, the number of molecules in the simulations ranged from 3400 up to 55000.
In simulations with polymer, two 10.5~kDa PEG chains of the formula
HO--[--CH$_2$--CH$_2$--O--]$_n$--H,  $n=239$, were placed into the simulation box along with water and ions.

The supporting square underlay in the simulations without periodic boundary conditions was realized by a square grid of Lennard-Jones (LJ) atoms with fixed positions. The spacing between neighbours in the grid was 0.14142~nm. The contact angle depends on the mutual interaction between the droplet molecules and the underlay. We fixed, therefore, the parameters to the values corresponding to a weakly hydrophobic material to provide a sufficiently firm attachment while preventing the undesired spreading of liquid over the underlay surface. Furthermore, in order to prevent the droplet from moving over the edge of the underlay, a repulsive rim made of sites with the $\sim r^{-12}$ repulsive potential was added. The parameters of the LJ potential $C_{12}r^{-12}-C_6r^{-6}$ were, respectively, $C_{12}=1.234\times10^{-6}$~kJ\,mol$^{-1}$\,nm$^{12}$ and $C_6=5.852\times10^{-4}$~kJ\,mol$^{-1}$\,nm$^6$ for the underlay, and $C_{12}=4.937\times 10^{-6}$~kJ\,mol$^{-1}$\,nm$^{12}$ for the rim. For the interaction of the underlay with the liquid, geometric mean combining rules were applied to both constants (GROMOS default).

The intensity of the applied electric field varied within the range from 0.01 to 2~V\,nm$^{-1}$. The use of a field of such strength is consistent with simulation studies on water reported in the literature~\cite{Cramer,Kusalik,34}. In real electrospinning experiments, the applied voltage is, typically, between 10 and 30~kV. The exact strength around the apex of the Taylor cone depends on the electrode separation and the geometry of the apex. A rough estimate gives values between 1 and 100~V\,nm$^{-1}$ which also provides support for the strength of the field used in the simulations.

Before the field was applied, all systems were preequilibrated in an $NVT$ ensemble at 298.15~K using the stochastic velocity rescaling method of Bussi et al.~\cite{vrescale} with correlation time 0.1~ps. In simulations involving jet formation, no thermostat was in use after the field had been turned on, while in bridge stability studies, a thermostat was applied during the entire simulation. Ions were excluded from velocity scaling to avoid interference with their drift.

\section{Results and discussion}

\begin{figure}[!b]
  \begin{center}
    \includegraphics[width=0.9\textwidth]{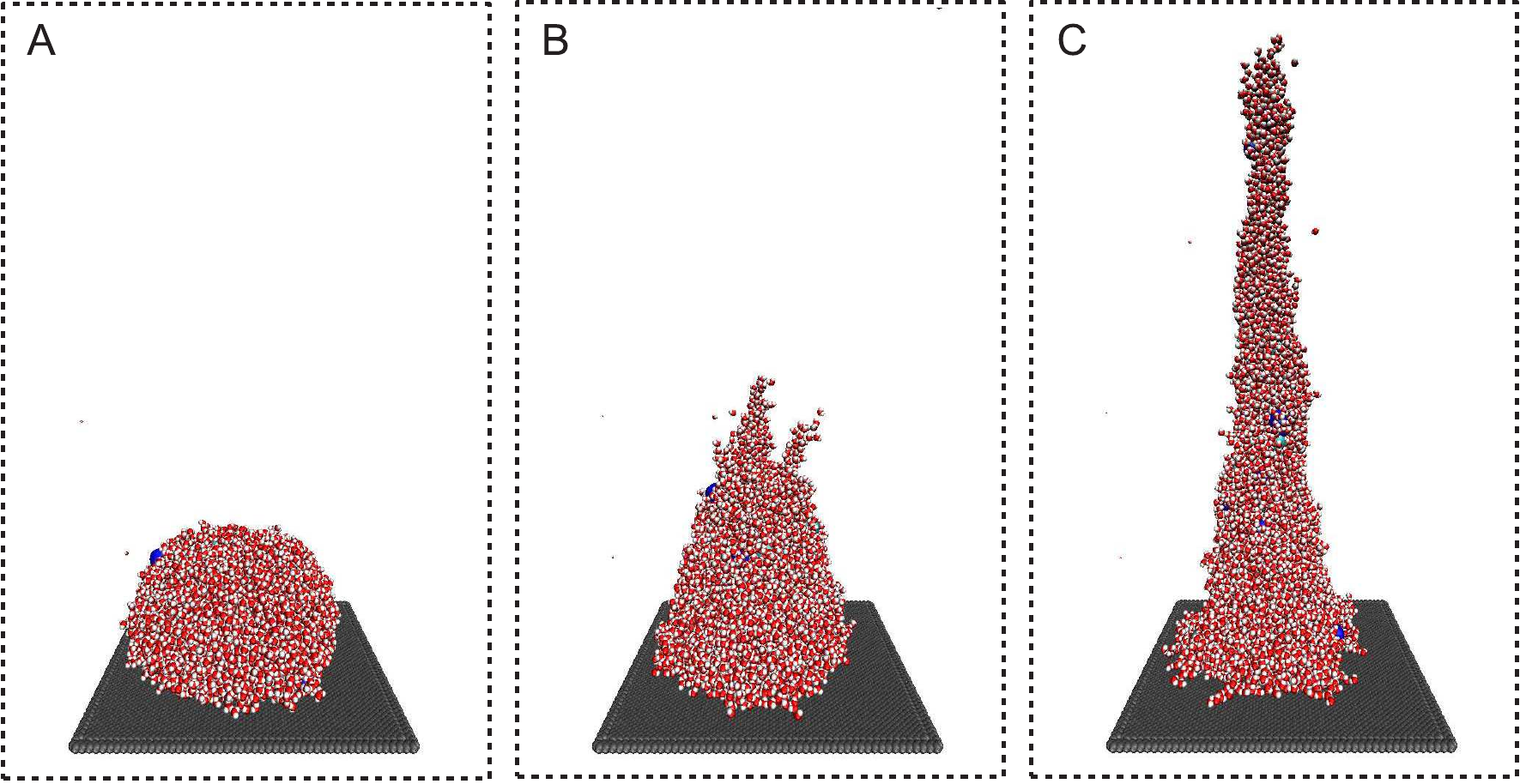}
  \end{center}
  \caption{(Color online) Snapshots showing the development of the jet (column structure) from a dilute solution of NaCl: (A) the initial droplet on an underlay; (B) transformation of the droplet into a cone structure; (C) establishing of the jet. The strength of the external electric field was 1.5~V\,nm$^{-1}$ and the droplet contained 18 NaCl ion pairs and 5954 molecules of water; snapshots (B) and (C) were taken, resp., 27~ps and 49~ps after the external electric field was switched on.}
  \label{fig_dilutejet}
\end{figure}

A typical development of the jet observed in simulations with dilute solutions of NaCl is shown in figure~\ref{fig_dilutejet}. Qualitatively the same behavior was also observed for pure water. Without the external field, the solution remains on the underlay in the form of a droplet, figure~\ref{fig_dilutejet}~(A). When the homogeneous external field in the direction perpendicular to the underlay is switched on, the droplet gradually develops into a cone, figure~\ref{fig_dilutejet}~(B), and then it forms a pillar resembling a jet in electrospinning. Apart from a small number of individual water molecules belonging to its vapor phase, the jet remained continuous. It means that no molecular clusters broken off the jet (which is typical of electrospraying) were observed.
Since the results obtained for pure water and a dilute electrolyte were qualitatively the same, one can deduce that ions, i.e., charge bearing particles, are not the driving force behind the origin of jetting. One can thus speculate that the building blocks of the jet are the dipoles (either permanent or induced) of molecules. That the dipoles do form a chain aligned along the external field has been confirmed and this is demonstrated below. Nonetheless, the internal structure of pure water may be more complex because even at a strong external field the hydrogen bonding network remains intact despite the alignment of a dipole as shown by Cramer et al.~\cite{Cramer}.

\begin{figure}[!t]
  \begin{center}
    \includegraphics[width=0.9\textwidth]{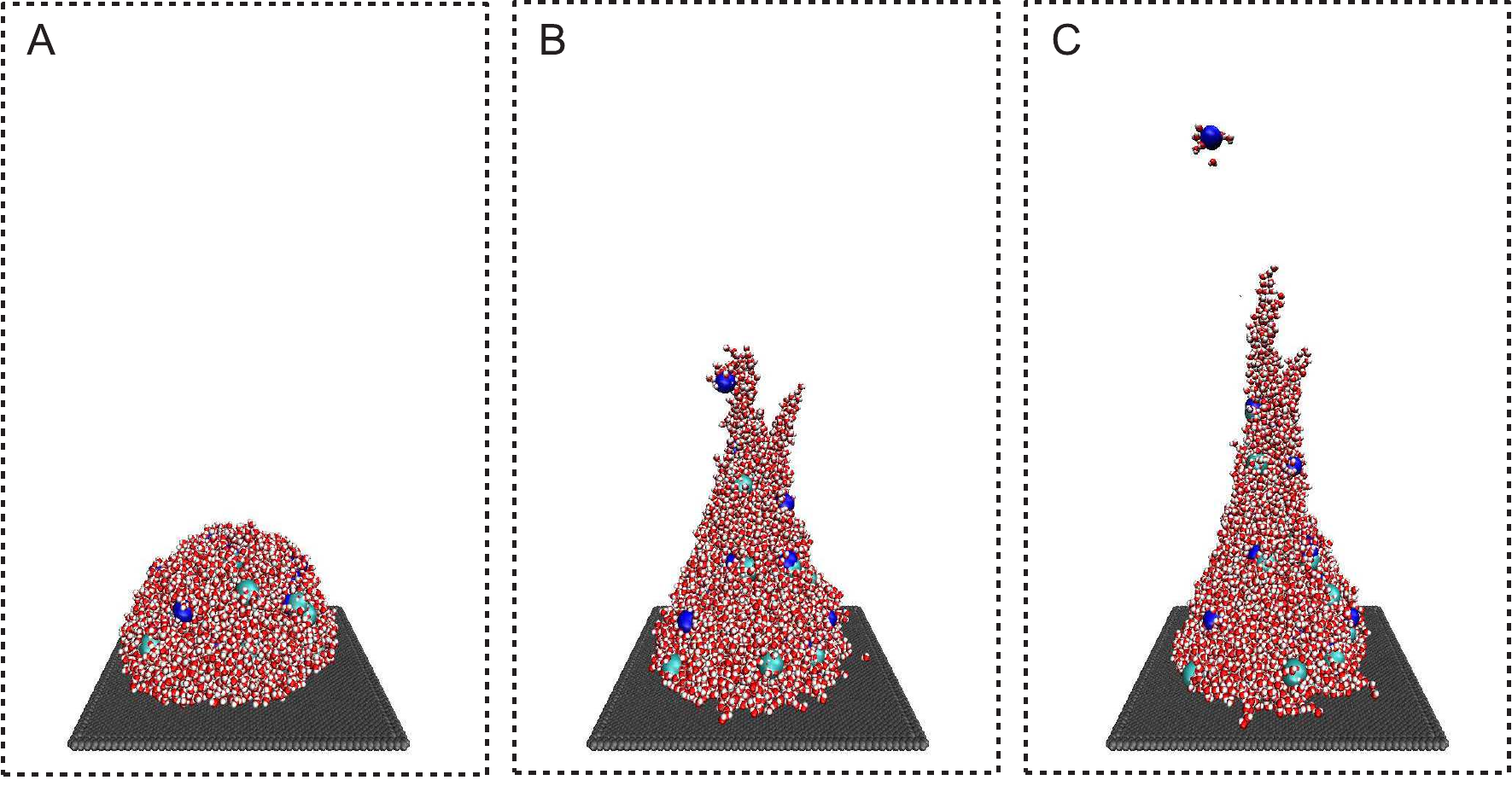}
  \end{center}
  \caption{(Color online) Snapshots showing the development of the jet (column structure) and its disruption from the concentrated solution of NaCl: (A) the initial droplet on an underlay; (B) transformation of the droplet into a cone structure; (C) disrupted jet with a hydrated ion cluster leaving the jet. The strength of the external electric field was 1.5~V\,nm$^{-1}$ and the droplet contained 36 NaCl ion pairs and 5954 molecules of water; snapshots (B) and (C) were taken, resp., 30~ps  and 34~ps after the external electric field was switched on. The small droplet torn off the jet (C) contained 13 molecules of water and one Na$^+$ ion.}
  \label{fig_brokenjet}
\end{figure}

\begin{figure}[!b]
  \begin{center}
    \includegraphics[width=0.9\textwidth]{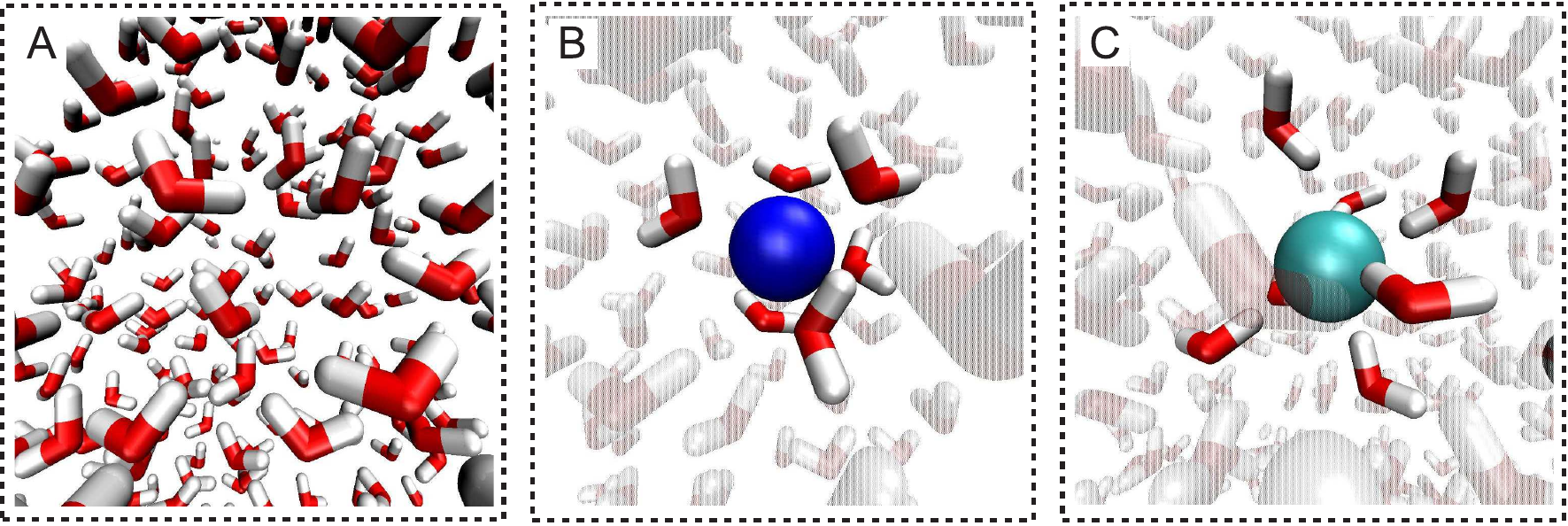}
  \end{center}
  \caption{(Color online) Snapshots showing the orientation of the water molecules in the jet: (A) in the ion-free region; (B) around Na$^+$ cation; (C) around Cl$^-$ anion. The strength of the external electric field was 1.5~V\,nm$^{-1}$. In order to highlight the molecules around the ions, water molecules whose oxygen-to-ion distances were greater than 0.3~nm (B) and 0.4~nm (C) are rendered in transparent colors.}
  \label{fig_orientation}
\end{figure}

In  figure~\ref{fig_brokenjet} we show the same sequence as above in the case of a more concentrated solution. At the beginning, the droplet follows the same evolution pattern as discussed above but at a later instant, instead of the development into a stable jet, small clusters formed by a central ion (hydrated ion) leave the bulk structure and accelerate away along the field lines of the external field. Although reality may be more complex, it is evident that the ions do disturb the structure observed in the case of pure fluids.
The interaction between the ions and water molecules (hydration) may overcome the effects of external field on water (polarization). This intuitive deduction is confirmed in figure~\ref{fig_orientation}. The strong polarization of water molecules and their alignment along the field in the jet is demonstrated in panel (A). Panel (B) reveals that the ion Na$^+$ makes the water molecules in its vicinity reorient with their oxygen atom to be closer to it. Similarly, the  Cl$^-$ ion has the same effect with the difference that now it is the hydrogen atom which is pulled closer to the ion (C). This reorientation of water molecules disturbs thus the original alignment of the water dipoles which in the case of sufficiently high concentration leads to disintegration of the column-like structure and spraying or collapse of the bridge.

\begin{figure}[!t]
  \begin{center}
    \includegraphics[width=0.9\textwidth]{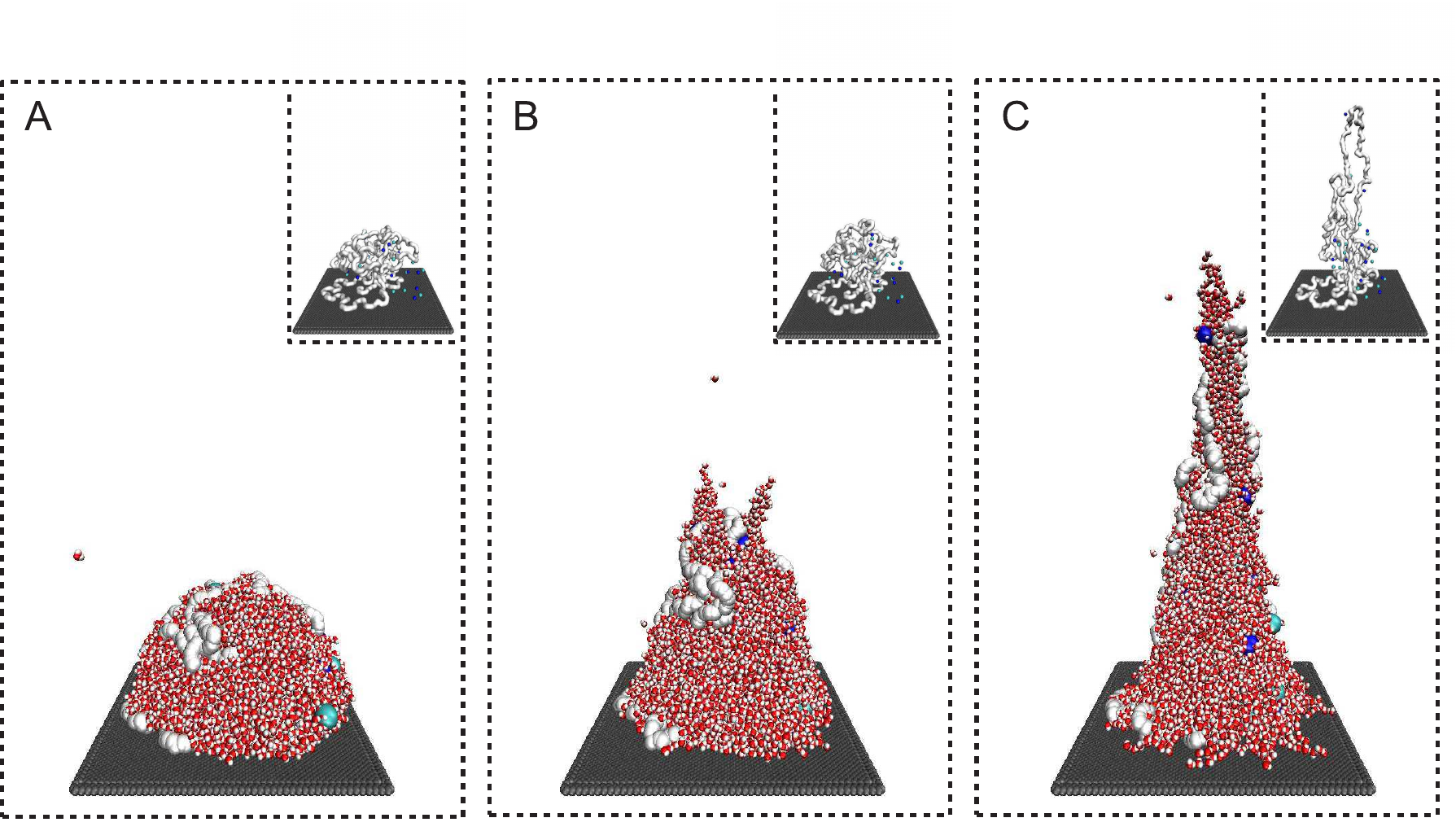}
  \end{center}
  \caption{(Color online) Snapshots showing the development of the jet from the solution of polyethylene glycol and NaCl in water: (A) the initial droplet on an underlay; (B) transformation of the droplet into a cone structure; (C) establishing of the jet. The strength of the external electric field was 1.5~V\,nm$^{-1}$; the droplet contained 2 chains of polyethylene glycol, 18 NaCl ion pairs, and 5954 molecules of water; snapshots (B) and (C) were taken, resp., 25~ps and 50~ps after the external electric field was switched on. In order to expose the polymer conformation, the same configurations are shown in the inserts where the water molecules are not displayed.}
  \label{fig_polymer}
\end{figure}

In our feasibility study~\cite{JMN_1} we considered only pure water and electrolyte solutions. In real electrospinning, it is polymer which in the end forms the nanofiber. Another step towards real electrospinning must also involve polymer molecules; we have also performed, therefore, simulations on a polymer solution. It is known that the majority of simulations with polymer solutions are carried out either on lattice or considering the solvent only implicitly. This is not suitable for studying the molecular nature of the electrospinning process and there are two questions to be answered: (i) whether a picture of the process can be obtained within a reasonable CPU time using explicit solvent and (ii) how large the polymer molecule should be to make simulations feasible and yet realistic.
In figure~\ref{fig_polymer} we show jet development from the polymer solution which does not differ much from the scenarios discussed above. The polymer molecule changes its conformation along with deformation of the droplet, figure~\ref{fig_polymer}~(B). The polymer chain uncoils when it is forced to enter the jet, figure~\ref{fig_polymer}~(C). A detailed investigation of the changes of the polymer conformations in dependence on the polymer composition, its structure, the strength of the field, electrolyte concentration, etc., is beyond the scope of this study and deserves an independent investigation. Nonetheless, the presented results show that such a study with polymer solution is nowadays feasible.

\begin{figure}[th]
  \begin{center}
    \includegraphics[width=0.75\textwidth]{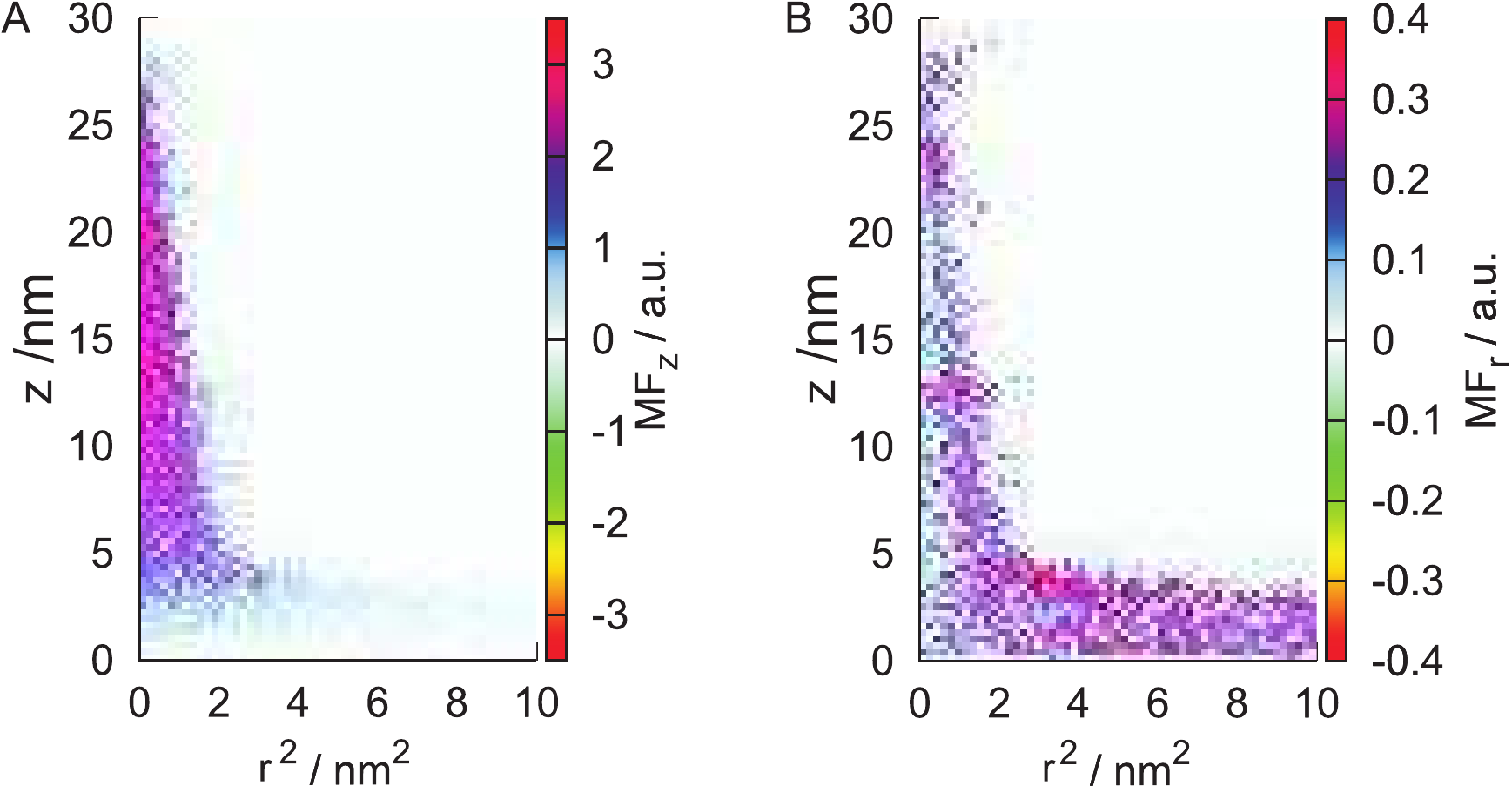}
  \end{center}
  \caption{(Color online) Average mass flow (MF) in the jet erupted from a flat layer of pure water at the external field of 2~V\,nm$^{-1}$. The profile is shown as a function of the squared distance $r^2$ from the axis of the jet and the distance $z$ from the supporting underlay: (A) flow in the direction along the external electric field; (B) flow in the direction towards the axis of the jet. The axis of the jet is considered parallel with the external electric field and going through the center of mass of the jet. The mass flow was averaged over the interval of 10~ps and it was expressed as a relative value in arbitrary units (the same in both panels).}
  \label{fig_flowprofile}
\end{figure}

In figure~\ref{fig_flowprofile} we show the results on the dynamics of the onset of jetting: the average mass flow (MF) in its dependence on the distance from the axis running through the jet's center-of-mass and parallel to the external field and also on the distance from the underlay. As it is seen, (A), deep in the bulk solution and far from the bottom of the jet, the effect of the external field on MF$_z$ is negligible. However, as soon as the solution enters the region of the jet, it sharply accelerates in the direction of the field. In figure~\ref{fig_flowprofile}~(B) we show the motion of the solution in the direction towards the jet's axis. Despite a large statistical noise, we see that on the surface of the solution in the vicinity of the base of the jet there is a region where the solution is more strongly drawn into the jet. We may thus gather that a relatively large part of the jet mass originates from the surface layer around the jet.

\begin{figure}[!b]
  \begin{center}
    \includegraphics[width=0.9\textwidth]{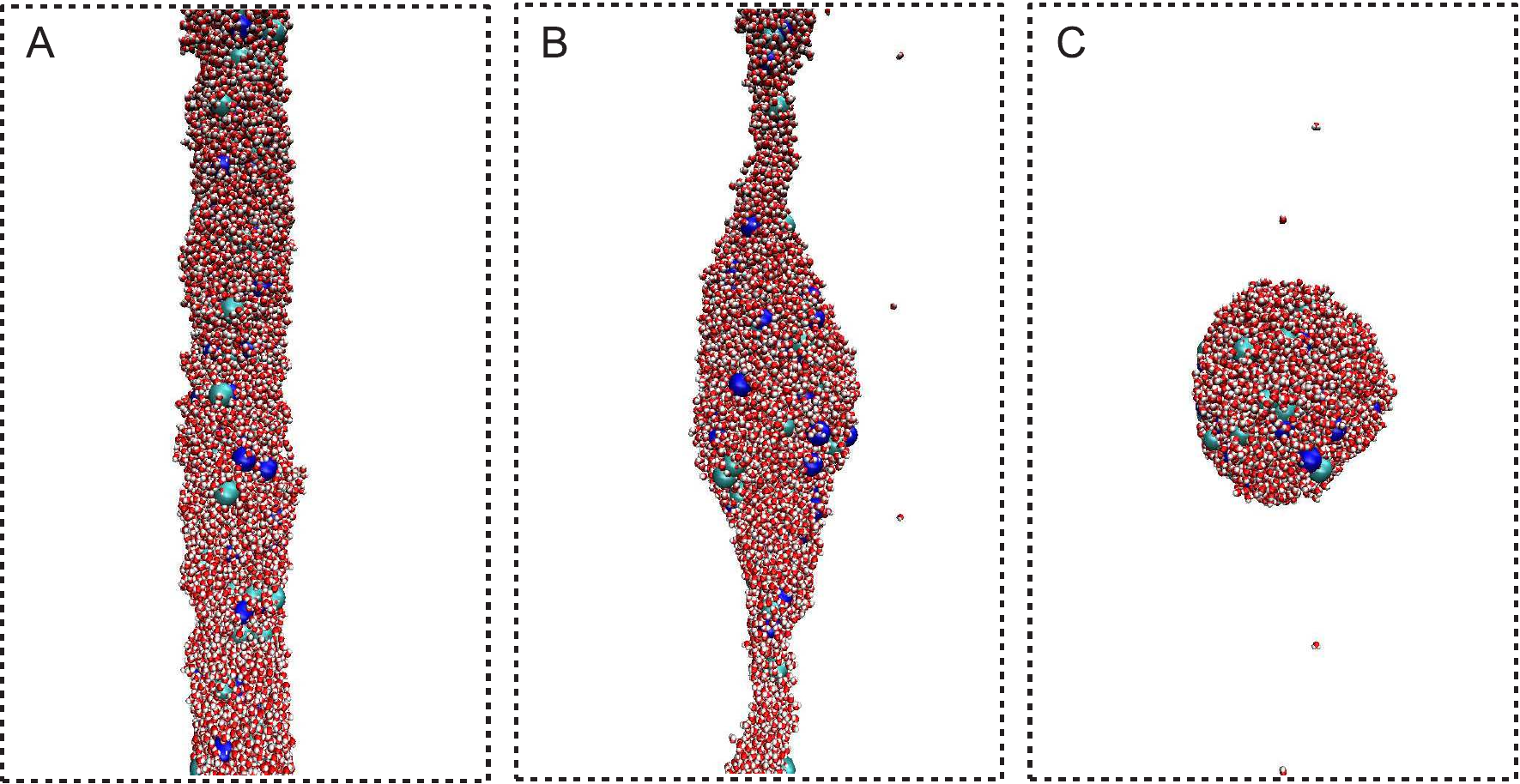}
  \end{center}
  \caption{(Color online) Sequence of snapshots showing the disintegration of a cylindrical nanobridge of aqueous solution of NaCl: (A) the established water nanobridge at the external field of 2~V\,nm$^{-1}$; (B) its gradual deformation at 0.1~V\,nm$^{-1}$; (C) its final collapse into a droplet. The solution contained 64 ion pairs and 6784 water molecules; snapshots (B) and (C) were taken, resp., 750~ps  and 900~ps after the external electric field was switched on.}
  \label{fig_bulb}
\end{figure}

All results commented on above correspond to situations where the external electric field is strong enough to initiate transformation of the liquid surface into a jet. If the field was suddenly weakened under the jetting threshold at any instant during the simulation, the jet would eventually vanish. However, one can use quite a different setup, where we anchor a liquid pillar on both ends, forming a liquid bridge. In contrast to a jet, being a dynamical phenomenon, the bridge can exhibit a thermodynamic (meta)stability (in the absence of drifting ions) and can exist under much a weaker field than that needed for its build-up from the liquid surface. An illustration of this phenomenon is given in figure~\ref{fig_bulb} showing snapshots from the simulation of the NaCl aqueous solution, which initially forms a column continuous in the $z$-direction, i.e., in the direction of the applied field, through periodic boundary conditions, see panel~(A). This setup elegantly avoids attaching the ends of the bridge to a particular material by simulating only the middle portion of the bridge self-anchored by its periodic images. The bridge was originally formed from a liquid slab by a strong field perpendicular to the liquid surface, in a way similar to all the above-described simulations. Subsequently, a column was subjected to a series of weakened fields, while the stability of the structure, i.e., whether a column remains continuous or breaks, was assessed. When the field was weak enough, increasingly prominent bulging appeared along the column, see figure~\ref{fig_bulb}~(B), resulting in the ultimate disintegration of the continuous liquid column into droplets, see panel~(C).

\begin{figure}[!t]
  \begin{center}
    \includegraphics[width=0.55\textwidth]{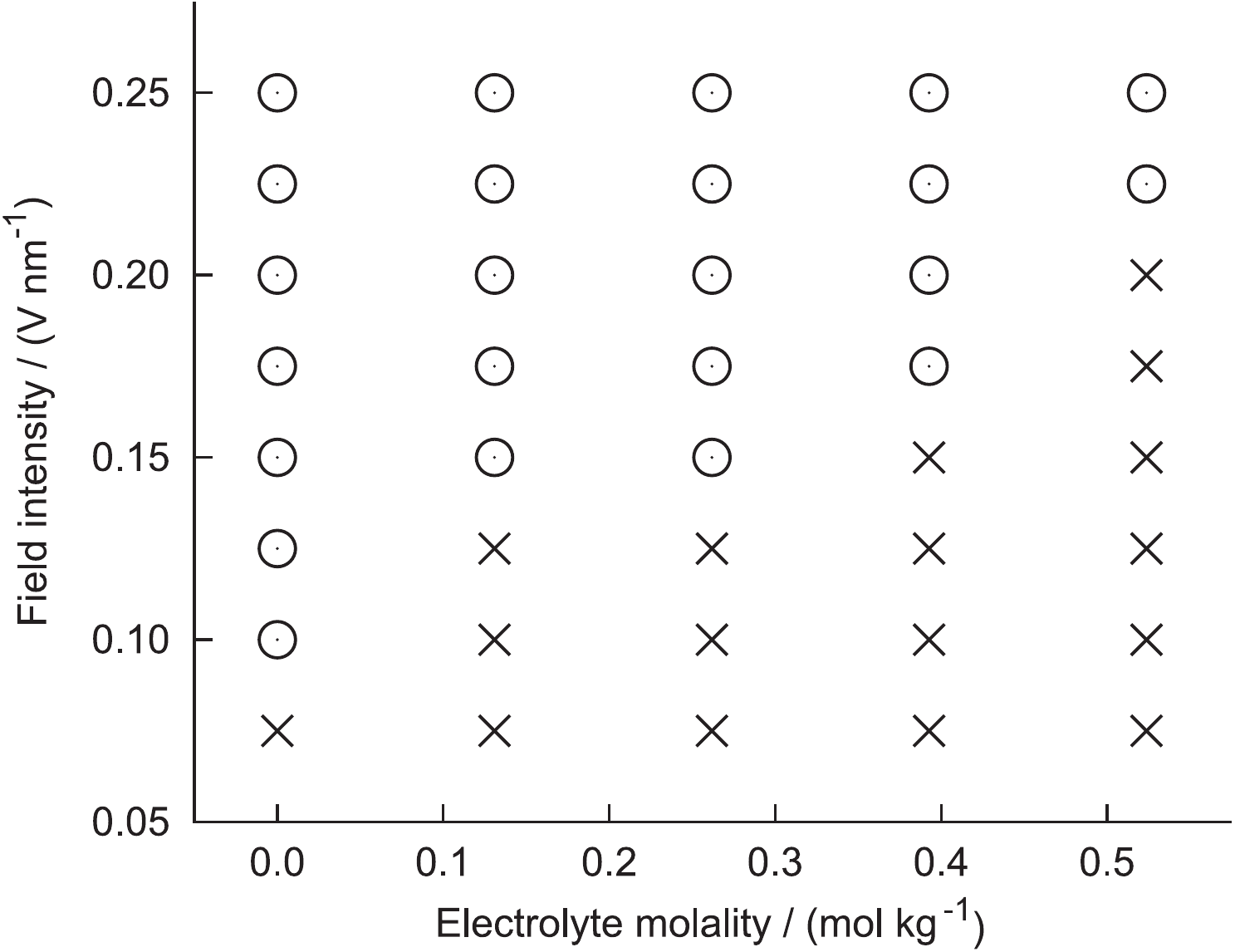}
  \end{center}
  \caption{Dependence of the electric field strength necessary for a stable bridge of aqueous NaCl solution to be formed on its concentration. All the initial bridges were 4~nm in diameter. Circles denote the stable bridge and crosses mark its breakup.}
  \label{fig_hysteresis}
\end{figure}

Figure~\ref{fig_hysteresis} maps the stability of the established bridge with respect to the strength of the field and the electrolyte concentration. In accord with the above-given explanation of the effect of ions on the jet stability (discussion pertaining to figure~\ref{fig_brokenjet}), one can see that as concentration increases, the critical intensity of the electric field needed to maintain a stable bridge also increases. It is to be expected that this behavior also depends on the dimensions of the bridge;  however, this was not investigated within the present study. All the bridges used in the stability assessment had diameters around 4~nm. The resulting field intensities are substantially larger than those usually applied in experiments with the FWB of macroscopic scales. One can thus expect that the critical field strength will decrease with increasing bridge diameter. However, a proper molecular study of this trend will require further extensive simulations.

\section{Conclusions}

Our study has shown that it is possible to model complex field-induced processes at liquid surfaces, such as the formation of the electrospinning jet or the collapse of the water bridge, at the atomistic level by molecular simulation. We have shown how such simulations can be performed and have given several examples of the microscopic phenomena which are experimentally inaccessible whilst being important for proper understanding of the behavior of aqueous solutions in electric fields. In contrast to intuitive views on the mechanism of the origin of jetting and bridging phenomena, our simulations show that the essential building blocks of the liquid column are the water molecules, and that this is qualitatively independent of the presence of ions in solutions. The ions, such as Na$^+$ and Cl$^-$, destabilize both the growing jet and the water bridge. The primary mechanism leading to the destabilization is the disruption of the water orientational polarization in the ions' vicinity. The stability thus decreases with growing ionic concentration. As regards the dynamics of jetting, the simulations indicate that the particles are drawn into the jet preferentially from the liquid surface near the base of the jet and subsequently accelerate in the direction of the field. Each of the above conclusions, however, provokes new questions, which can be reliably answered only after further extensive simulations are performed as well as thorough analysis.

\clearpage

\section*{Acknowledgement}

This work was supported by the Czech Science Foundation (Grant No.~P208/12/0105). Access to computing facilities provided under the program `Projects of Large Infrastructure for Research, Development, and Innovations' (LM2010005) is appreciated. The authors also would like to thank Professor Ronald G.~Larson (University of Michigan) and the members of his group, especially Mr.~Kyle Huston, for providing the parameter files for the PEG force field.


\clearpage

\ukrainianpart

\title{Застосування молекулярного моделювання: погляд на явища формування рідинних містків і струменів}
\author[I. Nezbeda \textsl{et al.}]
 {
 І. Незбеда\refaddr{PU,UCHP},
 Ї. Їрсак\refaddr{PU},
 Ф. Моучка\refaddr{PU},
 У.Р. Сміт\refaddr{UoG}}
 \addresses{
 \addr{PU}
 Факультет природничих наук, Університет Я.Е. Пуркіне,
 Усті над Лабом, Чеська республіка
 \addr{UCHP} Інститут фундаментальних основ хімічних процесів,
 Академія наук, Прага, Чеська республіка
 \addr{UoG} Факультет математики і статистики, Університет Гуелфа, Гуелф, Канада
 }

\makeukrtitle

\begin{abstract}

Методом молекулярної динаміки  виконано моделювання  чистої води, водного розчину хлориду натрію і полімерних розчинів,
 що піддавалися дії сильних зовнішніх електричних полів з метою отримання молекулярних характеристик структурного
 відгуку на дію цих полів. Щоб встановити молекулярні процеси, які ведуть до формування рідинних містків і струменів
 (при виробництві нановолокон), використано декілька методологій моделювання. Встановлено, що у структурах усталеного
 наномасштабу, молекули утворюють ланцюжок з дипольними моментами, орієнтованими паралельно до прикладеного поля
 по всьому об'єму зразка. Присутність іонів може внести збурення в цю структуру, що приведе до її повної
 дезінтеграції на краплини;
встановлено залежність порогового поля, необхідного для стабілізації стовпця рідини, від концентрації.
Окрім цього, спостерігались конформаційні зміни полімера в процесі формування струменя.
\keywords електрообертання, місток плаваючої рідини, рідини у полі, поверхні водного розчину, полімерний розчин, молекулярна динаміка

\end{abstract}

\end{document}